\documentclass[12pt,a4paper]{article}
\usepackage{amsmath}
\usepackage{amssymb}
\usepackage[nosort]{cite}
\usepackage[hyperref,bulletsep]{collect}
\def\gfxon{\usepackage[final]{graphicx}}

\gfxon

\makeatletter
\@addtoreset{equation}{section}

\makeatother

\makeatletter
\let\old@makecaption=\@makecaption
\def\@makecaption{\small\old@makecaption}
\makeatother

\makeatletter
\let\old@startsection=\@startsection
\renewcommand{\@startsection}[6]{\old@startsection{#1}{#2}{#3}{#4}{#5}{#6\mathversion{bold}}}
\makeatother

\let\oldPhi=\Phi
\let\oldPsi=\Psi
\let\oldGamma=\Gamma
\let\oldSigma=\Sigma
\renewcommand{\Phi}{\mathnormal{\oldPhi}}
\renewcommand{\Psi}{\mathnormal{\oldPsi}}
\renewcommand{\Gamma}{\mathnormal{\oldGamma}}
\renewcommand{\Sigma}{\mathnormal{\oldSigma}}

\newcommand{\hypref}[2]{\ifx\href\asklfhas #2\else\href{#1}{#2}\fi}

\newcommand{\tabref}[1]{Tab.~\ref{#1}}

\newcommand{\atopfrac}[2]{\genfrac{}{}{0pt}{}{#1}{#2}}
\newcommand{\sfrac}[2]{{\textstyle\frac{#1}{#2}}}
\newcommand{\half}{\sfrac{1}{2}}

\newcommand{\pint}{\makebox[0pt][l]{\hspace{3.4pt}$-$}\int}

\newcommand{\order}[1]{\mathcal{O}(#1)}

\newcommand{\superN}{\mathcal{N}}
\newcommand{\gym}{g_{\scriptscriptstyle\mathrm{YM}}}

\newcommand{\Tr}{\mathop{\mathrm{Tr}}}

\newcommand{\ellK}{{\rm K}}

\newcommand{\ellE}{{\rm E}}

\newcommand{\lrbrk}[1]{\left(#1\right)}
\newcommand{\bigbrk}[1]{\bigl(#1\bigr)}

\newcommand{\nn}{\nonumber}
\newcommand{\nln}{\nonumber\\}

\newcommand{\nlnum}{\\&&\mathord{}}

\newcommand{\earel}[1]{\mathrel{}&#1&\mathrel{}}
\newcommand{\eq}{\earel{=}}
\newenvironment{myeqnarray}{\arraycolsep0pt\begin{eqnarray}}{\end{eqnarray}\ignorespacesafterend}
\newenvironment{myeqnarray*}{\arraycolsep0pt\begin{eqnarray*}}{\end{eqnarray*}\ignorespacesafterend}

\def\[{\begin{equation}}
\def\]{\end{equation}}
\def\<{\begin{myeqnarray}}
\def\>{\end{myeqnarray}}

\newcommand{\adss}{$AdS_5 \times S^5$}
\newcommand{\td}{\tilde}
\newcommand{\sql}{\sqrt{\lambda}\ }
\newcommand{\vp}{\varphi}

\newcommand{\ep}{\epsilon}

\newcommand{\lra}{\leftrightarrow}
\newcommand{\ta}{\tau}

\newcommand{\wup}{{\rm w}}
\newcommand{\vo}{\omega} 
\newcommand{\tkappa}{\kappa} 

\newcommand{\cE}{\mathcal{E}}
\newcommand{\cS}{\mathcal{S}}
\newcommand{\cJ}{\mathcal{J}}

\ifx\href\asklfhas\newcommand{\href}[2]{#2}\fi
\newcommand{\arxivno}[1]{\href{http://arxiv.org/abs/#1}{#1}}

\begin{document}


\thispagestyle{empty}
\begin{flushright}\footnotesize
\texttt{\arxivno{hep-th/0308117}}\\
\texttt{AEI 2003-068}\\
\end{flushright}
\vspace{1cm}

\renewcommand{\thefootnote}{\fnsymbol{footnote}}
\setcounter{footnote}{0}

\begin{center}
{\Large\textbf{\mathversion{bold}Precision Spectroscopy of AdS/CFT }\par}
\vspace{1cm}

\textsc{
N. Beisert$^a$, 
S. Frolov$^{b,}$\footnote{Also at Steklov Mathematical Institute, Moscow.}, 
M. Staudacher$^a$
and A.A. Tseytlin$^{b,c,}$\footnote{Also at Lebedev Physics Institute, Moscow.}}
\vspace{5mm}

\textit{$^{a}$ Max-Planck-Institut f\"ur Gravitationsphysik, Albert-Einstein-Institut\\
Am M\"uhlenberg 1, D-14476 Golm, Germany}
\vspace{2mm}

\textit{$^{b}$ Department of Physics, The Ohio State University\\
 Columbus, OH 43210, USA}
\vspace{2mm}

\textit{$^{c}$ Blackett Laboratory, Imperial College\\
London,  SW7 2BZ, U.K.}
\vspace{3mm}

\texttt{nbeisert,matthias@aei.mpg.de}\\
\texttt{frolov,tseytlin@mps.ohio-state.edu}\par\vspace{1cm}

\vfill

\textbf{Abstract}\vspace{5mm}

\begin{minipage}{13.7cm}
We extend recent remarkable progress in the comparison of
the dynamical energy spectrum of rotating closed 
strings in $AdS_5 \times S^5$
and the scaling weights of the corresponding non-near-BPS
operators in planar ${\cal N}=4$ supersymmetric  gauge theory. 
On the string side the computations
are feasible, using semiclassical methods, if angular momentum 
quantum numbers are large. This results in
a prediction of gauge theory anomalous dimensions 
to all orders in the `t Hooft coupling $\lambda$.
On the gauge side the direct computation of these
dimensions is feasible, using a recently discovered 
relation to integrable (super) spin chains, 
provided one considers the lowest order
in $\lambda$. This one-loop computation then 
predicts the small-tension limit of the string
spectrum  for all (i.e.~small or large) quantum numbers.
In the overlapping window of 
{\it large}  quantum numbers and {\it small} effective
string tension, the string theory and gauge theory results 
are found to  match in a mathematically highly non-trivial fashion.
In particular, we compare energies of states with
(i) two  large angular momenta in $S^5$, and (ii)
one large angular momentum in $AdS_5$  and $S^5$ each,
and show that the solutions are related by an 
analytic continuation. Finally, numerical evidence is presented on 
the gauge side that the agreement persists also  at higher 
(two) loop order.

\end{minipage}

\end{center}

\newpage
\setcounter{page}{1}
\renewcommand{\thefootnote}{\arabic{footnote}}
\setcounter{footnote}{0}


\section{Introduction}
\label{sec:intro}

It is believed that free type IIB 
superstring theory on the \adss\ background is exactly dual to planar 
${\cal N}=4$ supersymmetric $SU(N)$ quantum gauge theory 
\cite{Maldacena:1998re,Gubser:1998bc,Witten:1998qj}. Here 
\[\label{eq:gs}
 4 \pi g_s=\frac{\lambda}{N}=0 \quad \mbox{for} \quad N=\infty \ ,
\ \ \ \ \ \  \lambda=N \gym^2= {\rm fixed} \ , 
\]
and $\lambda$ is the 't Hooft coupling constant.
Since the exact quantization of string theory in this curved background is 
not yet understood,
most of the results on the string side of the duality 
obtained until a year and a half ago were in the
(classical) supergravity limit
of infinite string tension 
${ \frac{1}{2\pi \alpha'}} \rightarrow \infty$,
which corresponds, via the ``effective string tension'' 
identification
\[\label{eq:tension}
\sqrt{\lambda}=\frac{R^2}{\alpha'} \ , 
\] 
to the strong coupling limit
on the planar gauge theory side.
Since in string theory little could be done at finite $\alpha'$ and $R$,
while in gauge theory little could be done at finite $\lambda$, 
until recently the perception was 
that any dynamical test of the AdS/CFT correspondence 
should be very hard to perform.
A notable exception were 
some successful studies of four-point functions
(involving BPS operators on the gauge side and supergravity
correlators on the string side) where some dynamical modes appear
in intermediate channels 
\cite{Arutyunov:2000ku,Eden:2000bk,Arutyunov:2002fh}.
This situation has dramatically improved due to new ideas
and techniques on both sides of the correspondence, which were 
largely influenced by the seminal work of \cite{Berenstein:2002jq}
(which, in turn, was based on \cite{Blau:2001ne,Metsaev:2001bj}).
Very recent progress points towards the exciting 
prospect that the free \adss~string {\it alias} planar gauge theory
is integrable and thus might be exactly solvable.

On the string theory side, it was understood that
in the
case when some of the quantum numbers of the string states become
large, the \adss\ string 
sigma model
can be efficiently treated by semi-classical methods 
\cite{Gubser:2002tv,Frolov:2002av} (see also \cite{Tseytlin:2002ny}
and \cite{Russo:2002sr,Minahan:2002rc,Mandal:2002fs}).
It was then suggested  \cite{Frolov:2003qc,Frolov:2003tu}
that a 
 novel possibility for a quantitative  comparison with SYM theory  
 in non-BPS sectors appears 
when one considers classical solutions describing closed 
strings rotating in {\it several} 
directions in the product space $AdS_5 \times S^5$
with the metric $ds^2_{10}= (ds^2)_{AdS_5} + (ds^2)_{S^5}$
\<\label{eq:dam}
(ds^2)_{AdS_5}
\eq d \rho^2 - \cosh^2 \rho \ dt^2 + \sinh^2\rho \ (d \theta^2 +
 \cos^2 \theta \ d \phi^2_1 + \sin^2 \theta \ d\phi_2^2) 
\nln
(ds^2)_{S^5}
\eq d\gamma^2 + \cos^2\gamma\ d\varphi_3^2 +\sin^2\gamma\ (d\psi^2 +
\cos^2\psi\ d\varphi_1^2+ \sin^2\psi\ d\varphi_2^2) 
\>
Here $t$ is the global $AdS_5$ time, which, together with 
the 5 angles $(\phi_1,\phi_2;\varphi_1,\varphi_2,\varphi_3)$, 
correspond to the obvious ``linear'' isometries 
of the metric, i.e.~are related to the 3+3 Cartan generators
of the $SO(2,4) \times SO(6)$ bosonic isometry group.
Rotating strings can thus carry the 2+3 angular momentum charges (spins) 
$Q_i =(S_1,S_2; J_1,J_2,J_3)$, while $t$ is 
associated with the energy $E$.
Once such classical solutions representing string states with several 
charges are found \cite{Frolov:2002av,Frolov:2003qc,Frolov:2003tu,Frolov:2003xy,Arutyunov:2003uj},
one may evaluate the energy as a function of the spins and $\lambda$:
$E=E (Q_i,\lambda)$. 
A remarkable feature of string solutions in $AdS_5$ 
is that their energy grows, for large charges, 
linearly with the charges \cite{deVega:1996mv,Gubser:2002tv,Frolov:2002av}.
Corrections 
to subleading terms in the classical energy 
can then be computed using the standard semiclassical (inverse string 
tension) expansion \cite{Frolov:2002av,Frolov:2003tu}. 
For certain string states with large total spin $J=J_1+J_2+J_3$ on $S^5$ for which
\footnote{Here $Q_i\sim J$ means that $Q_i=\gamma_i J$, where $\gamma_i$ are arbitrary 
constants which can be numerically large, small or even zero.
Then we can set up a power counting scheme in $1/J$ and $\lambda/J^2$. 
While we keep all orders of $\lambda/J^2$, we systematically
drop terms of subleading orders in $1/J$.}
\[\label{cond}
J\gg 1\ ,\qquad \frac{\lambda}{J^2} \ll 1\ ,\qquad
Q_i\sim J\ , 
\]
it turns out that string sigma model loop corrections to the energy
are suppressed by powers of ${\frac{1}{J}}$ (see 
\cite{Frolov:2003tu,Frolov:2003xy,Arutyunov:2003uj}).
In these cases the leading, $\order{J}$, contribution to the 
energy is given already by the {\it classical} string expression, 
i.e.~one does not even need to quantize the string sigma model.
Furthermore, as follows from  the string action,
classically, all charges (including the energy) appear only in the
combination $Q_i/\sqrt{\lambda}$.
It was shown in \cite{Frolov:2003qc,Frolov:2003tu,Frolov:2003xy,Arutyunov:2003uj} that 
the classical energy $E$ admits an expansion 
in powers of the small parameter $\lambda/J^2$.
The upshot of the semiclassical string analysis is then
that in the limit \eqref{cond}
the string-state energy is given by 
\[\label{ene} 
E(Q_i,\lambda)=S+J\lrbrk{1+\frac{\lambda}{J^2}\,
\epsilon_1(Q_i/J)+\frac{\lambda^2}{J^4}\,\epsilon_2(Q_i/J)+
\ldots}+\order{J^0},
\]
where $\epsilon_n(Q_i/J)$ are some functions of the 
spin ratios $Q_i/J$, and $\order{J^0}$ stands for quantum 
string sigma model corrections.

Considering string states represented by the 
classical solutions 
 with several charges 
$(S_1,S_2; J_1,J_2,J_3)$ has the added
advantage that it helps significantly in identifying
 the corresponding
gauge theory operators.
 As is well known, ${\cal N}=4$ supersymmetric gauge theory is
superconformally invariant,  and the bosonic subgroup of the full 
superconformal group $PSU(2,2|4)$ is $SO(2,4) \times SO(6)$. 
The energy $E$ of a string state is expected to 
correspond to the scaling dimension $\Delta$ of the associated 
conformal operator on the gauge theory side:
\[\label{eq:basic}
E(Q_i, \lambda)_{{\rm string}}=
\Delta(Q_i, \lambda)_{{\rm gauge}} \ . 
\]
The dimension $\Delta$ is, in general,
a non-trivial function of the 't Hooft coupling 
$\lambda$. 
It is not yet known how to compute $\Delta$ exactly,
except for supersymmetric BPS operators (for which 
dimensions are
protected, i.e.~independent of $\lambda$) and for certain 
near-BPS operators with large charge $J_3$
\cite{Berenstein:2002jq,Santambrogio:2002sb}.
In principle, one can compute the dimension $\Delta$ perturbatively 
at small $\lambda$ as an eigenvalue of the matrix of 
anomalous dimensions. 
Obtaining and diagonalizing this matrix is a task where the complexity 
increases exponentially with the number 
$L$ of constituent fields.
At this point,  a comparison to string theory energies 
may appear  almost hopeless, 
since, on  the string side, 
the total $SO(6)$ charge $J=L$ is required to be large.

Fortunately, it was recently discovered that planar
$\superN=4$ SYM theory is \emph{integrable} at the one-loop level
\cite{Minahan:2002ve,Beisert:2003yb}. 
We can therefore make use of the Bethe ansatz for a
corresponding spin chain model  to obtain \emph{directly} the 
eigenvalues of the matrix of anomalous dimensions.
This observation proves to be especially useful in the 
(``thermodynamic'') limit $L\gg 1$, i.e.~for a very long spin chain, where
 the  algebraic
Bethe equations are approximated by integral equations. 
For a large number of fields $J=L$, the dimension 
$\Delta$ appears to have a loop expansion equivalent
 to the one in \eqref{ene},
\footnote{For the $\order{\lambda}$ term the $1/J$ dependence 
can be read off from the thermodynamic limit of the Bethe ansatz.
For higher-loops, 
 there are some numerical indications for this $1/J$ dependence,
but so far  there is no 
general proof (which, perhaps, may be given 
using maximal supersymmetry of the theory).}
\[\label{eq:gaugeexp}
\Delta(Q_i,\lambda)=S+J\lrbrk{
1 + \frac{\lambda}{J^2} \,\delta_1(Q_i/J) 
+ \frac{\lambda^2}{J^4} \,\delta_2(Q_i/J) + \ldots} + \order{J^0}\,.\]
Again, the coefficients $\delta_n$ are functions of the spin ratios $Q_i/J$.

The string semiclassical expression 
\eqref{ene}, while formally valid for $\sqrt \lambda \gg 1$,
is actually exact, since, as was mentioned above, 
all sigma model corrections are suppressed by~$\frac{1}{J}$. 
Assuming the conditions \eqref{cond} are satisfied, one should be able
to compare directly the 
classical  $\order{J}$ term 
in the  string energy \eqref{ene} 
to the 
$\order{J}$ scaling dimension in gauge theory \eqref{eq:gaugeexp}, 
and to show that 
\[ \label{ekk}
\epsilon_n(Q_i/J)=\delta_n(Q_i/J).\]
Such a comparison at $\order{\lambda/J^2}$ order 
was indeed successfully 
carried out in \cite{Beisert:2003xu,Frolov:2003xy,Arutyunov:2003uj}, where 
a spectacular agreement 
between the string theory and gauge theory 
results for the energy or dimension 
was found for several 
two-spin string states represented by 
circular and folded closed strings rotating in $S^5$.

The contents of the present  paper is the following.
In 
Section~\ref{sec:S5} we   review the results of the semi-classical
computation of the energy of folded strings rotating in
two planes on $S^5$, the ``$(J_1,J_2)$'' solution \cite{Frolov:2003xy}. 
We  also review the results of the Bethe ansatz
calculations of the anomalous dimensions of the corresponding
gauge theory operators 
\cite{Beisert:2003xu}. 
We  then present a full  analytic \emph{proof} that in the 
region of large quantum numbers the relevant terms 
in the string energy and  the gauge operator dimension match, 
i.e that  $\epsilon_1$ and $\delta_1$ 
are indeed the same as {\it functions}  of  the spin ratio. 
This goes beyond 
the previous ``experimental'' evidence of matching series
expansions. 
The central part 
 of the present paper is Section~\ref{sec:AdS5},
where we show that the ``$(J_1,J_2)$''
state represented by the string rotating in 
two planes in $S^5$ can be analytically
continued, in {\it both} string and gauge theory, to 
an  ``$(S,J)$'' state represented by a  string
rotating in just one plane in $S^5$, but having 
also  one large spin in $AdS_5$ \cite{Frolov:2002av}.
 On the gauge theory  side,  this requires the use of a 
recently constructed \cite{Beisert:2003yb} supersymmetric extension 
of the above Bethe ansatz.
As a result, we  find the agreement between the string 
theory and gauge theory 
 expressions of the energy/dimension 
also for the ``$(S,J)$'' solution. 
In Section~\ref{sec:lambda2} we 
study the  possibility to check  the matching \eqref{ekk} 
beyond the leading order $n=1$. 
Using the  expression \cite{Beisert:2003tq}  for the 
gauge theory two-loop dilatation operator,
we present numerical evidence
that the  matching
 between the string  theory and  gauge theory results in the case of 
  the $(J_1,J_2)$ state  extends to
at least to the $n=2$  (two-loop) level.
Section~\ref{sec:conclusions} contains some concluding remarks,
The Appendices
contain some general remarks and technical details.
In particular, in Appendix~\ref{app:B} we explain the relation 
between the  $(J_1,J_2)$  and  $(S,J)$ string solutions, 
and in Appendix~\ref{app:C} we discuss the solution 
of the Bethe ansatz system
of equations for the spin chain  which appeared in 
Section~\ref{sec:AdS5} in connection with the $(S,J)$ case. 
In Appendix~\ref{app:E} we compare circular strings
on $S^5$ with a different ``imaginary'' solution of the Bethe equations.
Finally, in Appendix~\ref{app:D} we consider the dependence of 
string energy on the ratio of two spins.

\section{Strings rotating on $S^5$}
\label{sec:S5}

Let us start 
with a discussion  of  a particular $(J_1,J_2)$ 
state  corresponding to a  {\it folded}  string rotating in 
two planes on the five-sphere. This folded solution 
should have minimal value 
of the energy for given values of the spins.
Our aim will be to  
 demonstrate the equivalence between the leading correction
to the 
classical string-theory  energy  and the 
one-loop gauge theory anomalous 
dimension at the {\it functional} level, 
i.e.~going beyond  particular expansions and limits 
considered previously in \cite{Beisert:2003xu,Frolov:2003xy}.
The solution in question \cite{Frolov:2003xy} has the following 
non-zero coordinates in \eqref{eq:dam}:
$t=\kappa \tau,\ \varphi_1= \wup_1 \tau, \ 
 \varphi_2= \wup_2 \tau,\ \gamma= {\frac{\pi}{2}} , \
 \psi= \psi(\sigma)$ and $\psi$ satisfies a 1-d sine-Gordon 
 equation in $\sigma$.
 The string is stretched in $\psi$
with the maximal value $\psi_0$
(we refer to Appendices~\ref{app:A} and \ref{app:B} for details 
on the string solutions).
 The classical energy  and 
the angular momenta 
of the rotating string 
may be written as  
\[\label{eq:curly}E= \sqrt{\lambda}\,\mathcal{E}, \qquad
J_i =  \sqrt{\lambda}\,\mathcal{J}_i.
\] 
Here and from now on, curly letters correspond to 
charges rescaled by the inverse effective string tension, 
$1/\sqrt{\lambda}$. 
The parameters of the solution $\kappa, \omega_1,\omega_2$ 
are related via the conformal gauge constraint and the closed string
periodicity condition (we shall consider single-fold solution). 
Solving these conditions one may express 
the energy as a function of the spins, 
$E(J_1,J_2,\lambda)
=\sqrt{\lambda}\,\mathcal{E}(\mathcal{J}_1,\mathcal{J}_2)$. 
The expression for the energy 
 can then be found as a parametric solution of the following 
 system of two transcendental equations 
(see Appendix~\ref{app:B})
\[\label{eq:StringS5}
\lrbrk{\frac{\mathcal{E}}{\ellK(x)}}^2
-\lrbrk{\frac{\mathcal{J}_1}{\ellE(x)}}^2=\frac{4}{\pi^2}\,x\ , 
\qquad
\lrbrk{\frac{\mathcal{J}_2}{\ellK(x)-\ellE(x)}}^2
-\lrbrk{\frac{\mathcal{J}_1}{\ellE(x)}}^2=\frac{4}{\pi^2}\ ,
\]
where the auxiliary parameter $x= \sin^2 \psi_0  $ is 
the modulus of the
elliptic integrals $\ellK(x)$ and $ \ellE(x)$ of the first and
second kind, respectively
(their standard definitions can be found in the appendices in 
\eqref{eli},\eqref{eq:conds2}).
Eliminating $x$, one finds the energy as 
a function of 
$J=J_1+J_2$, the ratio $J_2/J$ and the
string tension $\sqrt{\lambda}$. 

Assuming that $\mathcal{J}={\mathcal{J}_1}+{\mathcal{J}_2}$
is large, i.e.~that the condition  \eqref{cond}  is satisfied, 
one can expand the solution for the energy in powers of the total spin
$J=\sqrt{\lambda}\,\mathcal{J}$ (cf. \eqref{ene}) 
\[\label{eq:expa}
\mathcal{E}=
\mathcal{J}+\frac{\epsilon_1}{\mathcal{J}}
+\frac{\epsilon_2}{\mathcal{J}^3}+\ldots \ , 
\quad \mbox{i.e. }\quad
E=
J+\epsilon_1\,\frac{\lambda}{J}
+\epsilon_2\,\frac{\lambda^2}{J^3}+\ldots \ ,  
\]
where $\epsilon_n=\epsilon_n(J_2/J)$ are functions of
the spin ratio.  
It  is a non-trivial observation that the string 
energy admits \cite{Frolov:2003xy} 
such an expansion   which then looks like a
perturbative expansion in $\lambda$. Moreover, quantum string sigma 
model corrections to $E$ are suppressed if $J\gg 1$
\cite{Frolov:2003tu,Frolov:2003xy}.

Turning attention to the gauge theory side, the natural operators 
carrying the  same  $SO(6)$ charges $(J_1,J_2)$ 
are of the general form
\[\label{eq:embryo}
\Tr Z^{J_1} \Phi^{J_2}  + \ldots \ , 
\]
where $Z$ and $\Phi$ are two of the three complex scalars of the
${\cal N}=4$ super YM  model. The dots indicate that one has to consider 
all possible orderings of $J_1$ fields $Z$ and $J_2$ fields $\Phi$
inside the trace: only very specific linear combinations of these
composite fields possess a definite scaling dimension $\Delta$, 
i.e.~are eigenoperators of the anomalous dimension matrix.  
These particular two-spin scalar operators   do
not mix with  any  other local operators that contain 
other  types of factors (fermions, field
strengths, derivatives) \cite{Beisert:2003tq}.
The    relation \eqref{eq:basic}
between the string state energy and dimension of the corresponding 
gauge theory  operator  then predicts
that there should exist  an operator%
\footnote{Since the folded string 
solution happens to have  lowest energy for given charges, the
corresponding  operator 
should  have the lowest dimension in this class of operators.}
of the form
\eqref{eq:embryo} whose exact%
\footnote{By exact we mean 
not only all-order in $\lambda$ 
 but also non-perturbative:
one does not expect instanton effects to be relevant
 in the planar gauge theory.} 
scaling dimension is given, for large $J$, by the solution of 
eqs.\eqref{eq:StringS5}. That means, in particular, that 
 eqs.\eqref{eq:StringS5} derived  from  {\it classical} 
 string theory  predict the anomalous dimension
of this operator to any order in  perturbation theory in $\lambda$!

Can we test this highly non-trivial prediction by a direct one-loop
computation in the gauge theory? 
In the case where $J$ is large,  
doing this from scratch by 
Feynman diagram techniques is a formidable task
due to the
large number of possible field orderings
(one needs to
diagonalize the  anomalous dimension matrix  whose size grows 
exponentially with $J$). 
What helps is the crucial  observation of ref. \cite{Minahan:2002ve}
that the one-loop anomalous dimension matrix 
for the   operators of the two-scalar type
\eqref{eq:embryo}  can be related to a Hamiltonian 
of an integrable Heisenberg spin chain 
(XXX$_{+1/2}$ model), i.e.~its eigenvalues  
can be found by solving the Bethe ansatz equations of the spin chain. 
The upshot of the Bethe ansatz procedure \cite{Beisert:2003xu} 
is that the system of equations 
diagonalizing the one-loop anomalous dimension matrix 
(for any,  small or large,  values of $J_1,J_2$) 
is given by
\[\label{eq:Bethe1}
\left( \frac{u_j+i/2}{u_j-i/2} \right)^J=
\prod_{\textstyle\atopfrac{k=1}{k\neq j}}^{J_2}
\frac{u_j-u_k+i}{u_j-u_k-i}\,, \qquad
\delta_1=\frac{J}{8\pi^2}\sum_{j=1}^{J_2}
\frac{1}{u_j^2+{1/4}}\,, 
\]
where again $J=J_1+J_2$ (we assume $J_1 \geq  J_2$).
This is an algebraic system of equations involving the
auxiliary parameters $u_j$, the so-called Bethe roots.
We need to find the $J_2$ roots $u_j$
subject to the condition that no two roots $u_j,u_k$ coincide
and the further constraint
\[\label{eq:momco}
\prod_{j=1}^{J_2}\frac{u_j+i/2}{u_j-i/2}=1\ , 
\]
which ensures that no momentum flows around the cyclic trace.
This yields the one-loop planar anomalous 
dimensions $\lambda\frac{\delta_1}{J}$ for the 
operators \eqref{eq:embryo}.
Here, we will restrict consideration 
to symmetric solutions,
i.e.~if $u_j$ is a root then $-u_j$ must be a root as well, 
this automatically solves the momentum constraint \eqref{eq:momco}.%
\footnote{
For a highest weight state of the desired representation $[J_2,J_1-J_2,J_2]$ of $SO(6)$,
no roots at infinity are allowed.}

Let us now  pause and 
compare the string theory
system \eqref{eq:StringS5} 
for the classical energy  and the gauge theory system 
\eqref{eq:Bethe1} for the one-loop anomalous dimension.
 Both systems are parametric, i.e.~finding energy/dimension as a function of
 spins 
 involves elimination
of auxiliary parameters. 
The string result is 
valid for all $\lambda$, but
 restricted to large $J$, 
namely,  $J \gg \sqrt{\lambda} $ {\it and}  $J \gg 1$.
The gauge result is valid for all $J_1,J_2$, but restricted 
to lowest order in $\lambda$. 
Remarkably, there is a region of joint validity:
large charge $J$ \emph{and} first order in $\lambda$!

Extracting the leading-order or ``one-loop''
 term $\ep_1$ from the string-theory relations 
\eqref{eq:StringS5} is straightforward, as discussed 
in Appendix~\ref{app:B}. 
For large $\mathcal{J}=\mathcal{J}_1+\mathcal{J}_2$ one sets
$x=x_0+x_1/\mathcal{J}^2+\ldots$
and solves the resulting transcendental equation for $x_0$. 
One then finds the parametric solution 
for $\ep_1=\ep_1 ({\frac{J_2}{J}})$
\[\label{eq:stringembryo}
\epsilon_1 =
\frac{2}{\pi^2} \ellK(x_0)
\bigbrk{\ellE(x_0)-(1-x_0)\ellK(x_0)}\ , \qquad
\frac{J_2}{J}=1-\frac{\ellE(x_0)}{\ellK(x_0)} \ . 
\]
On the gauge side, 
one needs to do some work to extract
the lowest-energy  state solution  \cite{Beisert:2003xu}.%
\footnote{It is worth noting that the Bethe equations
give the energies or dimensions for all  states
with the same spins, while the minimal-energy string solution 
corresponds to the ground state only.
To select  a particular solution of the Bethe equations that 
should correspond to a specific (folded or circular, with 
extra oscillations or without) string solution  
requires a number of steps: First, one needs to make certain 
``topological'' assumptions about the distribution of roots,
which accumulate on lines presciently termed ``Bethe strings''.
The possible choices correspond to ``folded'' and ``circular'' strings.
Second, one takes the logarithm on both 
sides of the Bethe equations. The possible branches correspond
to the various winding modes of the string.} 
First, to be able to compare with string theory 
we need to consider  the  ``thermodynamic'' limit of large spins,  
i.e.~$J\gg 1$.  
The idea is then to assume a condensation of the Bethe roots into
``strings''
and thus to  convert the system of algebraic 
Bethe equations into
a continuum (integral)  equation. 
Making an  appropriate ansatz for the 
Bethe root distribution selects the ground state. Solving the
corresponding 
integral equation (see Appendix~\ref{app:C} for some  details) one finds again 
a system of two equations 
with energy as a   parametric solution
$\delta_1=\delta_1 ({\frac{J_2}{J}})$
\[\label{eq:gaugeembryo}
\delta_1=\frac{1 }{2 \pi^2} 
\ellK(q)\bigbrk{2\ellE(q)-(2-q)\ellK(q)} 
,\qquad
\frac{J_2}{J}
=\frac{1}{2}-\frac{1}{2\sqrt{1-q}}\,\frac{\ellE(q)}{\ellK(q)}  \ . 
\]
Here the modulus $q=1-\frac{a^2}{b^2}$ is related to the
endpoints $a,b$ of the ``strings'' of Bethe roots.
This system  looks  similar, but superficially not identical  to that in 
eq.\eqref{eq:stringembryo}. 
However, if we  relate  the
auxiliary parameters $x_0$ and $q$ by
\footnote{Note  that $b=a^\ast$
(cf.  Appendix~\ref{app:C}),  so 
$x_0$ is indeed positive. Let us note also that 
this relation between the size of the 
folded string ($\psi_0$)  and  the ``length''  of the 
Bethe ``strings'' ($q$)  suggests that 
a transformation between the two integrable models 
-- string sigma model (Neumann system or 1-d sine-Gordon 
system that follows from it)  and the  spin chain --
should involve some kind of a Fourier transform
(Bethe roots are inversely related to effective  1-d momenta).} 
\[\label{eq:map}
\sin^2\psi_0\Big|_{\mathcal{J}=\infty}  =
 x_0=-\frac{(1-\sqrt{1-q})^2}{4\sqrt{1-q}}=-\frac{(a-b)^2}{4ab}, 
\]
one can show, using the elliptic integral modular transform 
relations 
\[\label{eq:Modular ansform}
\ellK(x_0)=(1-q)^{1/4}\ellK(q),\quad\ \ 
\ellE(x_0)=\half (1-q)^{-1/4} \ellE(q)+\half (1-q)^{1/4}\ellK(q) \ , 
\]
that the systems  \eqref{eq:stringembryo} and \eqref{eq:gaugeembryo}
are, in fact, exactly  the same. 
As a result, their solutions $\epsilon_1({\frac{J_2}{J}})$ and 
$\delta_1({\frac{J_2}{J}})$ do become  identical!
  
 We have thus demonstrated the equivalence 
 between the string theory and gauge theory results 
 for a particular two-spin part of the spectrum 
at the full functional level. 
Previously the equality $\epsilon_1=
\delta_1$  was checked 
 \cite{Beisert:2003xu,Frolov:2003xy} 
only for the first few 
terms in an expansion around special values 
of ${\frac{J_2}{J}}$.

\section{Strings rotating on $AdS_5$ and $S^5$}
\label{sec:AdS5}

Recently, it was shown in \cite{Beisert:2003yb} that 
the complete one-loop planar dilatation operator of $\superN=4$ SYM 
\cite{Beisert:2003jj} is integrable.%
\footnote{Integrability is related to Yangians. 
A Yangian structure in the bosonic coset sigma model 
was recently shown \cite{Bena:2003wd} to have a generalization to 
(classical) 
supercoset sigma model of \cite{Metsaev:1998it}.
 Very recently \cite{Dolan:2003uh}, this structure was
``mapped'' to planar gauge theory. 
Possibly,  this line of thought will
lead to a deeper understanding of the 
matching of energies/anomalous dimensions.} 
To diagonalize 
\emph{any} matrix of anomalous dimensions, 
the corresponding Bethe ansatz 
was  written down in
\cite{Beisert:2003yb}.
This enables one to access a much wider class of states
and perform similar comparisons between gauge theory and semiclassical 
string theory. 

Here we will present a first interesting example of such a novel test:
we shall consider the case of only one non-vanishing angular momentum 
in $S^5$ ($J=J_3$), but also  one non-zero 
spin in $AdS_5$ ($S=S_1$).
This situation is clearly different from the one
discussed  in the last section. However, on the string side, 
the two scenarios are,  in fact,  mathematically closely related, as
we will explain below (see also Appendix~\ref{app:B}). Is this also true on the gauge side?
There the relevant local operators
carrying the  same  $SO(2,4) \times SO(6)$
charges $(S,J)$ have the following generic form
\[\label{eq:baby}
\Tr {\cal D}^S Z^J+\ldots \ , 
\]
where ${\cal D}={\cal D}_1  + i{\cal D}_2$ is a complex 
combination of covariant derivatives 
(see \cite{Frolov:2003qc} for a related discussion).
Can we also treat these operators by a Bethe ansatz?
The integrability property of anomalous dimensions of 
 similar operators was 
recently discussed in the literature \cite{Belitsky:2003ys}.
In \cite{Beisert:2003yb} the precise spin chain interpretation 
of these operators was proved to lead to 
an integrable XXX$_{-1/2}$ Heisenberg chain, and
the corresponding Bethe ansatz was obtained.
Here, the derivatives $\mathcal{D}$ do not represent 
sites of the spin chain, in contradistinction to the 
fields $\Phi$ of \eqref{eq:embryo}.%
\footnote{In fact, the analogy goes the opposite way: $\Phi$ should be viewed as an
equivalent of $\mathcal{D}Z$, whereas 
$\mathcal{D}^{2,3,\ldots}Z$ are absent in the spin $+\half$ chain.}
In other words,  each site can now 
a priori (i.e.~if $S$ is sufficiently large) be 
in \emph{infinitely} many spin states $(\mathcal{D}^k Z)$, 
as compared to only \emph{two}, $(Z,\Phi)$, in \eqref{eq:embryo}.
Under this identification  $S$ plays the role of the {\it number} 
of excitations and $J$ equals the number of spin sites,
i.e.~the {\it length} of the chain.
The Bethe ansatz  equations for the one-loop 
anomalous dimensions then read 
(we use $\tilde\delta$ and $\tilde\epsilon$ to distinguish
the $(S,J)$ solutions)
\[\label{eq:Bethe2}
\left( \frac{u_j-i/2}{u_j+i/2} \right)^J=
\prod_{\textstyle\atopfrac{k=1}{k\neq j}}^{S}
\frac{u_j-u_k+i}{u_j-u_k-i}\,,
\qquad
\tilde\delta_1=\frac{J}{8\pi^2}\sum_{j=1}^{S}
\frac{1}{u_j^2+1/4}
\]
The similarity to the system of equations \eqref{eq:Bethe1} 
for the previous $(J_1,J_2)$ case, 
i.e.~for the  XXX$_{+1/2}$ spin chain,  is obvious. 
In fact, the   system  \eqref{eq:Bethe1} 
 becomes  formally equivalent to \eqref{eq:Bethe2} 
if we make the following replacements in 
\eqref{eq:Bethe1} 
\[\label{eq:gaugemap}
J \mapsto\  -J\ ,\qquad \ 
J_2\mapsto S\ ,\qquad\  \delta_1 (J_2,J) \mapsto -\tilde\delta_1(S,-J) \ .
\]
The large $J$ solution $\delta_1(J_2/J)$ of \eqref{eq:gaugeembryo}
can now be analytically continued to the regime $J_2/J<0$
where it gives the correct energy 
$\tilde\delta_1 (S/J)$ for $S/J>0$.
In fact,  the solution of \eqref{eq:gaugeembryo} was first derived 
\cite{Beisert:2003xu} by 
assuming that $J_2/J< 0$ and then analytically 
continued to $J_2/J> 0$! For further details see
Appendix~\ref{app:C}, where we also review the solution and
present some further results that were not included
in the paper \cite{Beisert:2003xu}.

Let us now turn to the rotating folded $(S,J)$ string 
solution \cite{Frolov:2002av} which would be expected to
correspond to the just discussed gauge theory operators
\eqref{eq:baby}. 
This rotating  string is stretched in the radial direction of $AdS_5$  
while its center of mass rotates in $S^5$,
it has the following 
non-zero coordinates in \eqref{eq:dam}:
$t= \kappa \tau,\ \rho=\rho(\sigma), \ 
\phi_1=  \omega_1 \tau, \ 
\varphi_3= \wup_3 \tau$ 
(see Appendices~\ref{app:A} and \ref{app:B} for details). 
Now the energy $E$ and the spin $S$  can be viewed as two  ``charges'' 
in $AdS_5$  while $J$ -- as the charge in $S^5$.
This is clearly reminiscent of the previous  example
where we had two charges $(J_1,J_2)$ in $S^5$   and one charge $(E)$
in $AdS_5$, 
and we have just found evidence on the gauge side that
one should actually expect the two solutions to be related by an analytic 
continuation. 
Indeed, as explained in Appendices~\ref{app:A} and \ref{app:B},
 a beautiful way to see
this connection on the string side stems from
 the close relation between the 
$AdS_5$ and $S^5$ metrics in  \eqref{eq:dam}.

On the level of the final expressions for the string 
charges  the relation is as follows.
The analogue of the parametric system of equations for the energy
in the $(J_1,J_2)$ case here is easily  found, using the relations 
in \cite{Frolov:2002av} (see Appendix~\ref{app:B}). 
We have  again   $E= \sql {\mathcal{E}}, \ S_1\equiv S= \sql {\mathcal{S}}, \
J_3\equiv J=\sql {\mathcal{J}}$,  where ${\mathcal{E}}, {\mathcal{S}}, 
{\mathcal{J}}$ depend only on the classical parameters 
$\kappa,  \omega_1 , \wup_3 $ and satisfy
\[\label{eq:StringAdS5}
\lrbrk{\frac{\mathcal{J}}{\ellK(x)}}^2
-\lrbrk{\frac{\mathcal{E}}{\ellE(x)}}^2=\frac{4}{\pi^2}\,x\ ,
\qquad
\lrbrk{\frac{\mathcal{S}}{\ellK(x)-\ellE(x)}}^2
-\lrbrk{\frac{\mathcal{J}}{\ellK(x)}}^2=\frac{4}{\pi^2}\,(1-x) \ . 
\]
The parameter $x= - \sinh^2 \rho_0$
here is  {\it negative} definite
 for a physical folded rotating string  solution. 
The system  \eqref{eq:StringAdS5}   becomes formally  equivalent to 
the one in \eqref{eq:StringS5} after  the 
following replacements done in \eqref{eq:StringS5}
(we choose the same signs of the charges as in Appendix~\ref{app:B})
\[\label{eq:stringmap}
\mathcal{E}\mapsto - \mathcal{J},\qquad\ \ 
\mathcal{J}_1\mapsto - \mathcal{E},\qquad\ \ 
\mathcal{J}_2\mapsto \mathcal{S} \ , 
\]
{\it and} after the analytic continuation from $x >0$ to $x<0$ in  the elliptic
integrals.
A formal relation between the solutions of the two systems 
\eqref{eq:StringS5} and \eqref{eq:StringAdS5}
is then 
\[\label{eq:InvertSol}
\mathcal{E}(\mathcal{J}_1,\mathcal{J}_2)\mapsto \ - 
\mathcal{J}(-\mathcal{E}, \mathcal{S})\ .
\]
In general, this does not imply a direct  relation between the energy expressions 
in the two cases: 
one needs to perform the analytic continuation and also to invert
the expression for $\mathcal{J}$. 
However, in the limit of large 
charges  (the limit we are interested in)
one can show that
the {\it leading} correction $\tilde\epsilon_1$ to the energy of the
 $(S,J)$ solution 
\[\label{eq:expa2}
\mathcal{E}=
\mathcal{S}+\mathcal{J}+\frac{\tilde\epsilon_1}{\mathcal{J}}
+\frac{\tilde\epsilon_2}{\mathcal{J}^3}+\ldots \ , 
\quad \mbox{i.e. }\quad
E=
S+J+\tilde\epsilon_1\,\frac{\lambda}{J}
+\tilde\epsilon_2\,\frac{\lambda^2}{J^3}+\ldots \ ,  
\]
is indeed  directly related to
  $\epsilon_1$
\eqref{eq:stringembryo} in the case  of the $(J_1,J_2)$ 
solution with  the replacements implied by  \eqref{eq:stringmap}
(see Appendix~\ref{app:B}).
In particular, to the leading order  in large-charge expansion 
one  has $J\equiv J_1 + J_2 \to - E +S \approx - J$ 
(where in the second  equality $J$ stands for $J_3$), 
so that 
$\tilde\epsilon_1 (-\frac{S}{J}) = - \epsilon_1 (\frac{J_2}{J}),$
i.e.~the two functions are related by 
$\tilde\epsilon_1 (-j) = - \epsilon_1 (j).$
Remarkably, this is 
the {\it same} as 
(the ``thermodynamic'' limit of) the 
 relation \eqref{eq:gaugemap} 
 found  above 
between the one-loop energy corrections 
on  the gauge theory side. 
This implies, in particular, that the correspondence 
between the string theory and gauge theory results 
for the leading terms in the energy/dimension 
holds also in the case of the $(S,J)$ states.

\section{Higher loop corrections}
\label{sec:lambda2}

Let us now comment on a generalization of the above 
results to higher orders 
in $\lambda$ (``higher loops'').  
First, let us note that on the string side, 
we have a complete expression for the energies to all orders in $\lambda$ 
which follows from the systems \eqref{eq:StringS5}
and \eqref{eq:StringAdS5}. 
In the interaction picture of perturbation theory, 
the only non-trivial system of equations is the one
determining the leading order contribution; 
all higher-loop terms can be expressed through 
the leading order modulus $x_0$. The 
two-loop energies $\epsilon_2$ for the $(J_1,J_2)$ case and 
$\tilde\epsilon_2$ for the $(S,J)$ case are given by 
(see Appendix~\ref{app:B})
\<\epsilon_2\eq\frac{2}{\pi^4}\, (\ellK(x_0))^3
\bigbrk{(1-2x_0)\ellE(x_0)-(1-x_0)^2\ellK(x_0)},
\nln
\tilde \epsilon_2\eq
-\frac{2}{\pi^4}\, (\ellK(x_0))^3
\bigbrk{\ellE(x_0)-(1-x_0^2)\ellK(x_0)}.
\>
As implied by  the relation 
\eqref{eq:InvertSol}, the two  expressions 
are not expected to (and do not)  look similar.

Given that integrability and the Bethe ansatz allow us to obtain 
the exact \emph{one-loop} energies for infinite length operators, while 
string theory gives us an all-loop prediction, it 
would be interesting to find \emph{higher-loop} energies in gauge theory 
to compare to string theory.
Although the integrability property of the dilatation operator 
acting on the states \eqref{eq:embryo} seems to be maintained 
(at least) at the two-loop level \cite{Beisert:2003tq},
the corresponding extension of the Bethe ansatz is not yet known.%
\footnote{In principal  agreement with the string theory result, 
one might express higher-loop energies in terms of the one-loop Bethe roots.
However, this would require calculating matrix elements of the higher-loop
dilatation operator between Bethe states --  presently a very 
non-trivial issue.}
Therefore, in order to find higher-loop 
anomalous dimensions of the operators \eqref{eq:embryo}
we have to rely on numerical methods of diagonalization of 
the matrix of anomalous dimensions.
For the states \eqref{eq:embryo}, this matrix is generated by the planar
dilatation operator \cite{Beisert:2003tq}
\<\label{dilop}
D(\lambda)\eq
J+
\frac{\lambda}{8\pi^2}\,\sum_{k=1}^J\bigbrk{1-P_{k,k+1}}
\nlnum\nonumber
+\frac{\lambda^2}{128\pi^4}\,\sum_{k=1}^J
\lrbrk{-4+6P_{k,k+1}-P_{k,k+1}P_{k+1,k+2}-P_{k+1,k+2}P_{k,k+1}}
+\order{\lambda^3},
\>
where $P_{k,k'}$ exchanges the fields at site $k$ and $k'$.
We shall consider the special case of $J_1=J_2=J/2$. 
For given $J$ we can collect all operators of the form
\eqref{eq:embryo}
and act on them with the dilatation operator
\eqref{dilop} neglecting all non-planar terms.
We can then diagonalize the matrix of anomalous dimensions and find 
the lowest eigenvalue of a state in the 
representation $[J/2,0,J/2]$, see \tabref{tab:SYM}.

\begin{table}\centering
$\begin{array}{|c|cccc|}\hline
J&\delta_1&\delta_2&\delta_3&\delta_3'\\\hline
\phantom{0}4&+0.303964&-0.123192&+0.087373&+0.124819\\
\phantom{0}8&+0.328847&-0.155138&+0.085280&+0.075704\\
12&+0.337964&-0.175313&+0.126497&+0.109772\\
16&+0.342407&-0.185011&+0.149043&+0.128745\\\hline
\infty&+0.356\ldots&-0.215\ldots&+0.212\ldots&+0.181\ldots\\\hline
\epsilon_n&+0.356016&-0.212347&+0.181347&+0.181347\\\hline
\end{array}$
\caption{SYM ground state energies for $J_1=J_2$. 
We show the one-loop and two-loop 
results as well as the three-loop conjectures.
The value at $J=\infty$ is obtained by extrapolating the
values for $J=8,12,16$ and 
$\epsilon_n$ represents the predictions from string theory.}
\label{tab:SYM}
\end{table}

The numerical one-loop results for finite $J$ are already reasonably close to 
the string theory prediction. As proposed in 
\cite{Beisert:2003xu}, we can improve the
results by extrapolating to $J=\infty$.
This is done fitting to the first two terms in the  series expansion 
in $1/J$ 
\[\label{fitfunc}
\delta_{n}=a_n+{\frac{b_n}{J}}+\ldots 
\]
As was demonstrated in \cite{Beisert:2003xu}, 
there are two distinct sequences of states for even and odd values of $J_2$.
The even values approach reasonably fast to the desired energies. 
Here we use $J=8,12,16$ to extrapolate to $J=\infty$.

The extrapolated values of the dimensions at 
one-loop and two-loop orders \cite{Beisert:2003tq}
are found to be about $1\%$ off the string theory prediction. 
These results agree very well and we can clearly confirm
that the correspondence works at $\order{\lambda^2}$!
For the three-loop conjecture of \cite{Beisert:2003tq} (see
also \cite{Beisert:2003jb}) the results are somewhat inconclusive.
On the one hand, using the vertex that was constructed assuming that 
integrability holds at $\order{\lambda^3}$, we get an extrapolation, $\delta_3$, 
which is $17\%$ off the string prediction. On the other hand,
the vertex that was matched to near plane-wave string theory results
\cite{Callan:2003xr} gives an extrapolation, $\delta_3'$, 
that is only $2\%$ away.
Nevertheless, we expect the three-loop energies to converge rather 
slowly and the three values of $J$ used to extrapolate are clearly not 
sufficient: In the spin chain picture the three-loop interaction already
extends over four lattice sites, and finite size effects, due
to the relatively small chain lengths, become more pronounced.  
An indication for this is that the extrapolation is
still $50\%$ away from the input values. The $1/J^2$ terms which
were neglected are expected to have a much stronger influence on the
finite $J$ values as compared to $\order{\lambda}$ term.
Therefore, a $17\%$ mismatch seems reasonable and we can
neither confirm nor rule out any of the conjectured three-loop vertices
\cite{Beisert:2003jb} (or the correspondence at $\order{\lambda^3}$).

\section{Conclusions and Outlook}
\label{sec:conclusions}

In this paper we demonstrated that 
spectroscopy is becoming a
very precise and versatile tool for establishing the validity of the
AdS/CFT duality conjecture on a quantitative, dynamical level.
Following the  suggestion of  \cite{Frolov:2003qc,Frolov:2003tu}
and 
extending the earlier break-through  work of 
\cite{Beisert:2003xu,Frolov:2003xy,Arutyunov:2003uj} 
we  have shown that in the non-BPS 
 sector of two  large 
charges, as in the near-BPS  BMN 
sector with  single large charge \cite{Berenstein:2002jq}, 
the  duality between the SYM theory and \adss\ string theory 
relates perturbative results on  both  sides of the correspondence
and thus can be tested using  existing tools.

It should be fairly 
evident that our derivation of mathematically highly
involved energy expressions,  such as 
eqs.\eqref{eq:stringembryo},\eqref{eq:gaugeembryo},
from both string theory and gauge theory
constitutes a ``physicist's proof'' of the correspondence.
We believe  that the present work is just 
the  beginning of a 
much wider unraveling of  dynamical details of the  AdS/CFT 
duality. At the end, we expect to gain much insight into
superstring theory on curved backgrounds, and into 
gauge theory at finite coupling.

Our work suggests a large number of further inquiries.
The precise interpretation of the circular versus
folded string solutions remains somewhat obscure in the Bethe ansatz 
picture. 
In particular, it would be important to understand the
analog of the string solution for $J_2>J/2$ in the Bethe ansatz
and thus complete the picture outlined in Appendix~\ref{app:D}.
Furthermore, it seems that the Bethe ansatz allows for
very complicated distributions of ``Bethe root strings'',
involving multi-cut solutions, the  role of which 
is unclear so far  on the  string theory side.

Another obvious problem is to extend the comparison to include $1/J$ terms
by computing (as in  \cite{Frolov:2002av,Frolov:2003tu})
the  1-loop string sigma model correction
to the $(J_1,J_2)$ string energy  
and comparing the result to the leading 
correction to the ``thermodynamic''
limit of the XXX$_{+1/2}$ Bethe system. 
It would be interesting also to compute energies
of excited string  states by expanding the superstring action 
near the ground-state two-spin $(J_1,J_2)$ solution.
In contrast to the BMN case \cite{Berenstein:2002jq}, 
here one expects (from experience with special circular
solutions \cite{Frolov:2003tu}) that there will be many
nearby states with the same charges and with energies differing from 
the ground state energy by order $\frac{1}{\cJ^2}= \frac{\lambda}{J^2}$ terms
 (these are of course negligible as  compared to similar terms in the
 classical ground-state  energy in the limit $J \gg 1$).

It should be relatively straightforward, if laborious to extend the analysis
to more than two spins. In string theory this has largely been accomplished
for three non-vanishing angular momenta on $S^5$ in
\cite{Arutyunov:2003uj}, but one could try to also include 
concurrently the two $AdS_5$ charges. For gauge theory,
the corresponding Bethe equations are known \cite{Beisert:2003yb},
but have not yet been analyzed in any generality.
Ideally, one would like to understand 
how to prove these equivalences directly, i.e.~without
actually solving the classical string sigma model equations
and the Bethe equations in the thermodynamic limit.

The biggest challenge clearly is to find out  how to extend
the calculational power on either side of the correspondence
in a way that would allow one  to derive results that are  {\it not} 
in the overlapping window of large quantum numbers
and small effective string tension.
 On  the string theory side,  this
would require to include   quantum 
(inverse string tension) corrections  in  the
Green-Schwarz supercoset sigma model of  \cite{Metsaev:1998it}.
For gauge theory, we would need to 
understand the proper extension of the Bethe ansatz
so as to make it applicable to all orders in Yang-Mills perturbation theory.
Maybe integrability will lead the way.

\subsection*{Acknowledgments}

We are grateful to J.~Russo and K.~Zarembo for useful discussions.
In particular, we thank G.~Arutyunov for many useful comments and
help with the discussion in Appendix~\ref{app:E}.
A.T.~is also grateful to the organizers of
Simons workshop in Mathematics and Physics at Stony Brook
for the hospitality at the workshop
during which this paper was completed.
The  work of S.F.~and A.T.~was supported by the DOE grant
DE-FG02-91ER40690. The work of A.T.~was also supported in part
by the  PPARC SPG 00613 and  INTAS  99-1590 grants and the Royal
Society  Wolfson award. N.B.~dankt der \emph{Studienstiftung des
deutschen Volkes} f\"ur die Unterst\"utzung durch ein 
Promotions\-f\"orderungsstipendium.


\appendix

\section{Rotating string solutions}
\label{app:A}

Let us  make  some general observations on 5-spin
  string  solutions  in \adss\ 
  pointing out some   relations between
different types of solutions via an analytic continuation.
The general rotating strings  carrying 
2+3 charges $(S_1,S_2; J_1,J_2,J_3)$ and the energy $E$ 
  are described by the following ansatz \cite{Frolov:2003qc}
  (see \eqref{eq:dam})
\[\label{anss}
\begin{array}{c}
t= \kappa \ta \ , \quad \phi_1 = \vo_1 \ta\ , \quad
\phi_2 = \vo_2 \ta\ ,  \quad
\vp_1= \wup_1 \ta \ , \ \ \vp_2= \wup_2 \ta \ ,\ \ \
 \vp_3= \wup_3 \ta \ , \\[6pt]
\rho(\sigma)= \rho(\sigma + 2 \pi)  \ , \quad
\theta(\sigma)= \theta(\sigma + 2 \pi)  \ ,\quad
\gamma(\sigma)=  \gamma(\sigma + 2 \pi) \ , \quad
 \psi(\sigma)=  \psi(\sigma + 2 \pi) .
\end{array} 
\]
Then the 3+3 obvious   integrals  of motion  are%
\footnote{As 
discussed in \cite{Frolov:2003qc,Arutyunov:2003uj}, all other 
$(S_{IJ},J_{MN})$ 
generators (conserved charges) 
of $SO(2,4) \times SO(6)$ except the Cartan ones 
$E= S_{05}, \ S_1= S_{12},\  S_2=S_{34}$, 
\ $J_1= J_{12}, \ J_2= J_{34},\  J_3=J_{56}$
should vanish in order for the rotating string solution to
represent a semiclassical string state carrying the
corresponding quantum numbers.}
\[\label{inti}\arraycolsep0pt
\begin{array}{rclcrcl}
\mathcal{S}_1
\earel
{\equiv}\displaystyle\frac{S_1}{\sqrt{\lambda}} = 
\vo_1 \int^{2 \pi}_0 \frac{ d \sigma}{2 \pi} \, \sinh^2 \rho\,  \cos^2 \theta\,,
 &\quad&\ 
\mathcal{J}_1
\earel
{\equiv}\displaystyle\frac{J_1}{\sqrt{\lambda}}=
\wup_1  \int^{2 \pi}_0 \frac{d \sigma}{2 \pi} \, \sin^2 \gamma \,  \cos^2 \psi \,,
\\[15pt]
\mathcal{S}_2
\earel
{\equiv}\displaystyle\frac{S_2}{\sqrt{\lambda}} = 
\vo_2  \int^{2 \pi}_0 \frac{d \sigma}{2 \pi} \, \sinh^2 \rho\,  \sin^2 \theta\, ,
 &\quad&\ 
\mathcal{J}_2
\earel
{\equiv}\displaystyle\frac{J_2}{\sqrt{\lambda}}=
\wup_2  \int^{2 \pi}_0  \frac{d \sigma }{2 \pi} \,  \sin^2 \gamma \,  \sin^2 \psi\, , 
\\[15pt]
\mathcal{E}
\earel
{\equiv}\displaystyle\frac{E}{\sqrt{\lambda}}= 
\kappa \int^{2 \pi}_0 \frac{d \sigma}{2 \pi} \, \cosh^2 \rho\, ,
 &\quad&\ 
\mathcal{J}_3
\earel
{\equiv}\displaystyle\frac{J_3}{\sqrt{\lambda}}=
\wup_3 \int^{2 \pi}_0  \frac{d \sigma}{2 \pi} \,  \cos^2 \gamma\, .  
\end{array}\]
They satisfy
\[
-\frac{\mathcal{S}_1}{\vo_1}
-\frac{\mathcal{S}_2}{\vo_2}
+\frac{\mathcal{E}}{\kappa}=1\,,\qquad 
 \frac{\mathcal{J}_1}{\wup_1} 
+\frac{\mathcal{J}_2}{\wup_2}
+\frac{\mathcal{J}_3}{\wup_3}=1
\, .\]
The second-order 
 equations for $(\rho,\theta)$  
\[\label{rt}
\begin{array}{c}
\rho'' -  \sinh \rho\, \cosh \rho \, 
(\kappa^2 + \tau'^2 - \vo_1^2 \cos^2 \tau - \vo^2_2 \sin^2 \tau )   =0 \,,
\\[6pt]
(\sinh^2 \rho \ \tau')'   -  
(\vo_1^2 - \vo^2_2) \sinh^2 \rho\,  \sin \tau \, \cos \tau  =0 \,,
\end{array}
\]
 and  $(\gamma,\psi)$ 
\[\label{rta}
\begin{array}{c}
\gamma'' -  \sin \gamma\, \cos \gamma\, 
(\wup_3^2 + \psi'^2 - \wup_1^2 \cos^2 \psi - \wup^2_2 \sin^2 \psi )   =0  \,,
\\[6pt]
(\sin^2 \gamma \ \psi')'   -  
(\wup_1^2 - \wup^2_2) \sin^2 \gamma \, \sin \psi \, \cos \psi  =0 \,,
\end{array}
\]
are decoupled from each other. 
As explained in \cite{Arutyunov:2003uj}, the resulting system of equations 
is completely integrable, being equivalent 
 to a combination 
of the two
Neumann dynamical systems. As a result, there are  2+2 
``hidden'' integrals of motion, reducing the 
general problem to solution 
of two independent 
systems of   two coupled  first-order  equations, 
with parameters related through  the conformal gauge constraint 
\<\label{conf}
 \rho'^2  - \kappa^2 \cosh^2 \rho   + \sinh^2\rho \ ( \theta'^2 +
 \vo_1^2 \cos^2 \theta  + \vo^2_2\sin^2 \theta ) \qquad
\nn\\
+ \  \gamma'^2 +  \wup_3^2 \cos^2\gamma  + \sin^2\gamma\ (\psi'^2 +
\wup_1^2 \cos^2\psi + \wup^2_2 \sin^2\psi) \eq 0 \ . 
\>
Let us now observe the following  symmetry of the above
system.
The two metrics in \eqref{eq:dam}
 are related  by the obvious 
  analytic continuation 
and change of the overall sign,  which is equivalent  in the
present rotational ansatz \eqref{anss} case   to
\[\label{con}
\rho \lra  i \gamma \ , \ \ \ \  \ \ \theta \lra \psi \ , \ 
\ \ \ \ \  \kappa \lra -\wup_3  \ ,\ \ \ \ \ \  
\vo_1 \lra  -\wup_1\ ,\ \ \ \ \ \  
\vo_2 \lra  -\wup_2 
\  .  \]
 This transformation  maps  the system 
\eqref{rt} into the system \eqref{rta} and 
also preserves the constraint \eqref{conf}. Thus it 
formally maps   solutions into solutions.
Under  \eqref{con}
the conserved charges \eqref{inti}
(or Cartan generators of $SO(2,4) \times SO(6)$)
transform as follows
\[\label{chaa}
S_1 \lra J_1 \ , \ \ \ \ \ 
S_2 \lra J_2 \ , \ \ \ \ \ 
E \lra  - J_3  \ . \]
We could, of course,  assume instead of \eqref{con} 
that $\vo_1,\vo_2 \lra  \wup_1,\wup_2$ but  then  
$S_1,S_2 \lra -J_1,-J_2 $.
Note that  the transformed solutions 
may not necessarily have a natural 
physical interpretation.
In order for some  two physical solutions 
to be related by this analytic continuation prescription 
at least one of them  should 
 have  a non-vanishing $J_3$ spin
(which transforms into the energy of the solution). 

One can   find  also other  transformations that 
map solutions into solutions by combining \eqref{con}
 with 
special (discrete)  $SO(2,4) \times SO(6)$ isometries 
that do not induce other components  of the 
rotation generators except  the above Cartan ones 
(e.g.,  interchanging the angular coordinates 
induces interchanging of the charges in \eqref{inti}, 
etc.).  
Below  we shall consider such an example.

\section{Relation between two-spin solutions}
\label{app:B}

Let us now  show that the  two 
previously known two-spin folded string 
solutions  are,  in fact,  related by 
the above analytic continuation.
Firstly, there is the ``$(S,J)$'' solution  \cite{Frolov:2002av} 
\[ \label{sj}  
 \kappa,\vo_1, \wup_3 \not=0 \ , \ \ \ \
\rho= \rho(\sigma) \ , \ \ \ \ \theta=0\ ,\ \ \ \
\gamma= 0 \ , \ \ \ \ \psi=0 \ ,  \]
where the string is stretched in
the radial direction $\rho$ of $AdS_5$ . 
It rotates ($\vo_1$) in $AdS_5$ about its center of mass 
which in turn moves ($\wup_3$) along a large circle of $S^5$.
The gauge constraint \eqref{con} and integrals of motion \eqref{inti} become 
\[\begin{array}{c}
  \wup_3^2  +  \rho'^2  - \kappa^2\, \cosh^2 \rho   + \vo^2_1\, \sinh^2\rho =0 \ , 
\ \ \ \ \ \ \ \cJ \equiv \cJ_3 =  \wup_3 \ , 
\\[6pt]\displaystyle
\cS\equiv \cS_1 =  \vo_1   \int^{2 \pi}_0 
{\frac{ d \sigma}{2 \pi}} \ \sinh^2 \rho\  
  \ , \  \ \ \ \ \ \ \ 
 \cE= \kappa  \int^{2 \pi}_0 {\frac{ d \sigma }{2 \pi}} \ \cosh^2 \rho  \ . 
\end{array} \]
Secondly, we have the ``$(J_1,J_2)$'' solution 
 \cite{Frolov:2003xy}
where the string 
located at the center of $AdS_5$
is stretched ($\psi$) along a great circle of $S^5$
and rotates ($\wup_2$)
about its center of mass which moves ($\wup_1$)
along an orthogonal great circle of $S^5$:
\[ \label{jj}  
  \kappa,\wup_1, \wup_2 \not=0 \ , \ \ \ \
\rho= 0 \ , \ \ \ \ \theta=0 \ ,\ \ \ \
\gamma= {\frac{ \pi}{2}} \ , \ \ \ \ \psi=\psi(\sigma)  \ .  \]
The gauge constraint \eqref{con} and integrals of motion 
\eqref{inti} are then
\[\begin{array}{c}
  -  \kappa^2   +  \psi'^2 +
\wup_1^2 \cos^2\psi + \wup^2_2 \sin^2\psi  =0 \ ,  
\\[6pt]\displaystyle
  \cE=  \kappa  \ ,\ \ \ \ \
\cJ_1=  \wup_1   \int^{2 \pi}_0 { \frac{d \sigma }{2 \pi}} \    \cos^2 \psi 
 \ , \  \ \ \ \ 
\cJ_2 =  \wup_2    \int^{2 \pi}_0 { \frac{d \sigma}{2 \pi}} \    \sin^2 \psi
  \ .  
\end{array}\]
Following the discussion in Appendix~\ref{app:A} we 
 conclude that these two solutions are related
by the following 
analytic continuation:
\footnote{
Note that here $  E-S-J \to E-( J_1 + J_2) $.
   Choosing  instead $\vo_1  \to  \wup_2$, 
$\kappa \to  \wup_1, \ \wup_3 \to \kappa $ we  would 
get  $   E \to J_1, \ S \to -  J_2 , \ J \to E$, 
so that $  E-S-J \to - E+   J_1+J_2$. 
}
\[\label{caa}
\begin{array}{c}
\rho \to  i \psi \ , \
\ \ \ \ \  \kappa\to  -\wup_1  \ ,\ \ \ \ \ \  
\vo_1  \to  -\wup_2 \  , \ \ \ \ \  \wup_3 \to - \kappa \ , 
\\[6pt]
   E \to -J_1 \ , \ \ \ \ \ \ \ 
S \to  J_2  \ , \ \ \ \  \ \ \ 
 J \to  - E \ .
\end{array}\]
We can, in fact,  directly relate the systems of 
equations  expressing the  closed string periodicity
condition and definitions of  the 
respective energies and spins and thus 
relating the  three  integrals of motion 
 in the two cases (see, respectively, 
 \cite{Frolov:2002av} and \cite{Frolov:2003xy}).
In the first case we get the following relations
(we introduce the parameter $x <0$ related to  $\eta$
in \cite{Frolov:2002av} by $\eta=-1/x$,\  $x= - \sinh^2 \rho_0$, where $\rho_0$
is the maximal value of the radial $AdS_5$ coordinate)
\<\label{pui}
\frac{\tkappa^2- \wup_3^2}{\tkappa^2-\vo_1^2} \eq - \sinh^2 \rho_0 \equiv x <0 
\ ,\qquad
{ \mathcal{E}}= \tkappa+\frac{ \tkappa}{\omega_1}\,
\mathcal{S},\qquad\ \ 
 \mathcal{J}=\wup_3\  ,
\nln
\sqrt{\tkappa^2-\wup_3^2}\eq
\frac{2\,\sqrt{-x}}{\pi}\,\ellK(x)\ ,
\qquad \ \ \ \ \ 
 {\mathcal{E}} 
=\frac{2\tkappa\,\sqrt{-x}}{\pi\sqrt{\tkappa^2-\wup_3^2}}\,\ellE(x)\  ,
\>
where 
$\ellK(x)$ and $\ellE(x)$ are the standard elliptic integrals
 (see Appendix~\ref{app:C}) 
related to the hypergeometric functions used  in \cite{Frolov:2002av} by 
\[ \label{eli} {}_2F_1(\half,\half;1,x)=
\frac{2}{\pi}\,\ellK(x)\ , \ \ \ \ \ \ \ 
{}_2F_1(-\half,\half;1;x)
=\frac{2}{\pi}\,\ellE(x)\ . 
\]
Solving  for $\omega_1$ and $\tkappa$ in terms of $\mathcal{J}$ and $x$
we find
\[\label{eq:StringOmKa}
\omega^2_1 =\mathcal{J}^2+\frac{4}{\pi^2}\,(1-x)(\ellK(x))^2,
\qquad
\tkappa^2=\mathcal{J}^2-\frac{4}{\pi^2}\,x\,(\ellK(x))^2 \ , 
\]
and then  finally  get the system of two equations 
\eqref{eq:StringAdS5} for the energy  given in the main text. 
The second of the two  equations in  \eqref{eq:StringAdS5} 
  determines $x$ in terms of $\mathcal{S}$ and
$\mathcal{J}$, while 
the first  one then gives the energy as a function of the spins.

Similarly, for the $(J_1,J_2)$ solution \eqref{jj} one finds from 
the expressions given in \cite{Frolov:2003xy} (we assume $\wup^2_2 > \wup^2_1$)
\<\label{kui}
\frac{\kappa^2 - \wup^2_1 }{\wup_2^2 - \wup_1^2}\eq \sin^2 \psi_0 \equiv x >0 \ , 
\qquad
1=\frac{\mathcal{J}_1}{\wup_1}+\frac{\mathcal{J}_2}{\wup_2},\qquad
\mathcal{E}=\kappa,
\nln
\mathcal{J}_1\eq 
\frac{2 \wup_1}{\pi \sqrt{\wup_2^2-\wup_1^2}}\,\ellE(x)\ , \qquad 
\sqrt{\wup_2^2-\wup_1^2}
=\frac{2}{\pi}\,\ellK(x)\ .
\>
Solving  for $\wup_1$, $\wup_2$ in terms of $\mathcal{J}_1$ and $x$ 
\[
\wup_1^2= \left( \frac{\ellK(x)}{\ellE(x)}\,\mathcal{J}_1\right)^2 \ ,
\qquad
\wup_2^2= \left( \frac{\ellK(x)}{\ellE(x)}\,\mathcal{J}_1\right)^2 
+\frac{4}{\pi^2}\,\ellK(x)^2 \ , 
\]
we finish with the system of the two equations determining 
$\mathcal{E}= \mathcal{E} ( \mathcal{J}_1, \mathcal{J}_2)$
given in \eqref{eq:StringS5}.
A manifestation of the analytic 
continuation relation between both two-spin solutions 
is then the  equivalence of the two systems 
\eqref{eq:StringS5}  and \eqref{eq:StringAdS5} under the 
substitution \eqref{eq:stringmap} 
(and a continuation from $x >0$ to $x<0$ in the  parameter space).

Depending on the region of the parameter space  
(or values of the integrals of motion) one finds different functional 
form of dependence of the energy on the two spins. 
We discuss some aspects of 
this dependence in Appendix~\ref{app:D} below. 
A direct comparison with gauge theory 
 we are interested in here is possible in the case 
 when  the two spins $S$ and $J$ are both 
  large compared to $\sql$,  i.e.~$ \cS \gg 1, \  \cJ \gg 1$.
The analogous  limit \cite{Frolov:2003xy} for the $(J_1,J_2)$ solution is 
when $\cJ_1 \gg 1 , \  \cJ_2 \gg 1$. 
In the two  cases we can then expand 
the energies, e.g., in powers of the total $S^5$ spin $J$.
This  amounts to an  expansion 
in powers of $\cJ\equiv \cJ_3$ in the $(S,J)$ case and in powers of 
$\cJ\equiv \cJ_1+\cJ_2$ in the $(J_1,J_2)$ case, respectively, 
\<\label{jjs}
E\eq S + J  + \ \frac{\lambda}{J}\ \td \ep_1 (S/J) + 
 \frac{\lambda^2}{J^3}\,  \td \ep_2 (S/J)  + \ldots \,,\qquad
  J\equiv J_3, \ S \gg \sql  \,,  
\nln
E\eq J  
+\frac{\lambda}{J}\, \ep_1 (J_2/J) + 
 \frac{\lambda}{J^3}\, \ep_2 (J_2/J)  + \ldots \,, \qquad J\equiv J_1 + J_2, J_2 \gg \sql \,, 
\>
where we introduced tildes on the correction functions $\ep_n$ 
in  the first  solution case. 
One may wonder if the coefficients $\td \ep_1$ and $ \ep_1$ 
in \eqref{jjs}  are related in some way, given 
that the two solutions are related by the analytic continuation. 
Applying  formally the substitution  \eqref{caa}   in 
\eqref{jjs} we get,  to the  leading order, 
\[\label{mak}
E= S + J  + \frac{\lambda}{ J}\, \td  \ep_1 (S/J) + \ldots
 \ \ \to \ \ 
-J_1 = J_2 - E  + \frac{ \lambda }{(-E)}\, \td   \ep_1 (J_2/(-E))  + \ldots \ . \]
Using that $ E = J_1 + J_2 +\ldots$ in the subleading term we finish then with
(where now $J\equiv J_1 + J_2$) 
\[\label{gott}
 E= J - \frac{\lambda }{J }\,  \td \ep_1 (-J_2/J)  + \ldots \ . \] 
Comparing this  to \eqref{jjs} we conclude that one should 
have a simple  relation between the leading-order (``one-loop'')
corrections for the energies of the two solutions:
\[ \label{rell}
  \td \ep_1 ( j ) =  -    \ep_1 (- j  ) \ .\]
As was noted in Section~\ref{sec:AdS5}, this 
 is indeed the relation that one finds on the gauge theory side
 \eqref{eq:gaugemap}.

 Let us now demonstrate  that  \eqref{rell}
follows also from the string-theory equations \eqref{pui} and 
\eqref{kui} or the systems 
\eqref{eq:StringAdS5} and  \eqref{eq:StringS5}. 
Expanding the parameter $x$ for large $\cJ$
 as (with $\cJ$ being 
 $\cJ_1 + \cJ_2$)
\[ \label{xex} 
x = x_0 + {\frac{ x_1 }{\cJ^2 }} + { \frac{x_2}{ \cJ^4} } + \ldots \ ,
 \] 
one finds that for the $(J_1,J_2)$ solution the leading value of the parameter 
$x_0$ is given by  the  solution of the transcendental equation 
\[\label{xxx}
\frac{\ellE(x_0)}{\ellK(x_0)}= 1-  \frac{J_2}{J} \ , \qquad\ \ \  
x_0= x_0 (J_2/J) \ .  \]
The rest of the expansion coefficients
in $x$ and the energy  are then determined by linear algebra, e.g., 
\<\nn
x_1\eq-\frac{4(1-x_0)x_0\bigbrk{\ellK(x_0)-\ellE(x_0)}\ellE(x_0)\ellK(x_0)^2}
{\pi^2\bigbrk{(1-x_0)\ellK(x_0)^2-2(1-x_0)\ellK(x_0)\ellE(x_0)+\ellE(x_0)^2}} \ , 
\\
\label{ee}
\epsilon_1 \eq \frac{2}{\pi^2} \,\ellK(x_0)
\lrbrk{\ellE(x_0)-(1-x_0)\ellK(x_0)}  \ , 
\\\nn
\epsilon_2  \eq \frac{2}{\pi^4} \,\ellK(x_0)^3
\bigbrk{(1-2x_0)\ellE(x_0)-(1-x_0)^2\ellK(x_0)}\ . 
\>
In the $(S,J)$ case,  using the same expansion \eqref{xex} for the corresponding 
parameter $x$  where 
now $\cJ=\cJ_3$ we find the following equation for  
 $x_0$
\[\label{yyy}
\frac{\ellE(x_0)}{\ellK(x_0)}=1+\frac{{S}}{{J}}\ , \ \ \ \ \ \ \ \ 
\ \ \   x_0 = x_0 (S/J) \ . 
\]
Solving this equation one finds other expansion coefficients in 
\eqref{xex} and \eqref{jjs}, e.g., 
\< \label{eee}
x_1\eq -\frac{4(1-x_0)^2x_0\bigbrk{\ellK(x_0)-\ellE(x_0)}\ellE(x_0)^2\ellK(x_0)}
{\pi^2\bigbrk{(1-x_0)\ellK(x_0)^2-2(1-x_0)\ellK(x_0)\ellE(x_0)+\ellE(x_0)^2}},
\\\nn
\td \epsilon_1\eq
-\frac{2}{\pi^2} \,
\ellK(x_0)\bigbrk{\ellE(x_0)-(1-x_0)\ellK(x_0)}\  ,
\\\nn
\td \epsilon_2\eq
-\frac{2}{\pi^4}\, \ellK(x_0)^3\bigbrk{\ellE(x_0)
-(1-x_0^2)\ellK(x_0)}\ .
\>
Comparing \eqref{xxx},\eqref{eee} to  \eqref{yyy},\eqref{ee}  and observing 
that to leading order \eqref{caa} implies $J_2 \to S, \ J\to - J$, 
we indeed  confirm the relation 
\eqref{rell}.

\section{Gauge theory details}
\label{app:C}

Here we will outline the solution of the 
Bethe ansatz system of equations
\eqref{eq:Bethe2} for the novel case of the XXX$_{-1/2}$
Heisenberg spin chain. We expect that the positions of the roots
are of order $\order{J}$, where $J$ is the length of our non-compact
magnetic chain, as explained in Section~\ref{sec:AdS5}.
We then take the logarithm of the equations \eqref{eq:Bethe2}
and obtain for large $J$ 
\[\label{eq:BetheRescale}
-\frac{J}{u_j}=2\pi n_j+2 \sum_{\textstyle\atopfrac{k=1}{k\neq
j}}^{S} \frac{1}{u_j-u_k}\ ,\ \ \ 
\qquad
\delta_1=\frac{J}{8\pi^2}\sum_{j=1}^{S}
\frac{1}{u_j^2}\ .
\]
The mode numbers $n_j$ enumerate the possible branches of the
logarithm. Excitingly, we see that these large $J$ equations
are almost identical to the ones found in \cite{Beisert:2003xu}
for the compact XXX$_{+1/2}$ chain (\emph{cf.} eq.(2.7) in
\cite{Beisert:2003xu}), 
except for a {\it minus sign} on the left 
hand side of the left equation in \eqref{eq:BetheRescale}.
It therefore does not come as a surprise that the solution
will be very similar to the previously considered  case.
The differences are,  however,  very interesting, and we will
briefly rederive the solution for the new case
\eqref{eq:BetheRescale}.

As in the case of the  XXX$_{+1/2}$ system, 
we shall start with  assuming  that
in the large $J$ limit 
 the Bethe roots accumulate on smooth
contours. It is reasonable, therefore, 
 to replace the discrete
root positions $u_j$ by a (rescaled) smooth continuum variable $u$
and introduce a density $\rho(u)$ describing the distribution
of the roots in the complex $u$-plane:
\[\label{eq:smooth}
\frac{u_j}{J} \rightarrow u
\qquad \qquad {\rm with} \qquad \qquad
\rho(u)=\frac{1}{J} \sum_{j=1}^S \delta \left(u-\frac{u_j}{J}\right).
\]
For the operators in eq.\eqref{eq:baby}
with one $AdS_5$ charge $S$ there are precisely $S$ roots,
and the density is normalized to the filling fraction
$\beta=S/J$, 
\[\label{eq:norm}
\int_{\mathcal{C}}du\, \rho(u)=\beta,
\]
where $\mathcal{C}$ is the support of the density, i.e.~the 
union of contours along which the roots are distributed. 
The Bethe equations \eqref{eq:BetheRescale} in the ``thermodynamic limit'' 
then conveniently turn into singular integral equations:
\[\label{eq:BetheCont}
\pint_{\mathcal{C}}\frac{dv\, \rho(v)\, u}{v-u}=\frac{1}{2}+\pi \,
n_{\mathcal{C}(u)} \,u,\qquad
\delta_1=\frac{1}{8\pi^2}\int_{\mathcal{C}}\frac{du\,\rho(u)}{u^2}
\]
where $n_{\mathcal{C}(u)}$ is the mode number at point $u$.
It is expected to be constant along each contour.
Here and in the following the slash through the integral sign
implies a principal part prescription.
In addition,  we have the
momentum conservation condition, resulting from the cyclic
boundary conditions of our chain:
\[\label{eq:Momentum}
\prod_{j=1}^S \frac{u_j+i/2}{u_j-i/2}=1\ , 
\qquad {\rm i.e.} \qquad
\int_{\mathcal{C}}\, \frac{du\, \rho(u)}{u}=0
\qquad {\rm and} \qquad
\int_{\mathcal{C}} du\, \rho(u)\, n_{\mathcal{C}(u)} = 0,
\]
where the last equation is a consistency condition derived from
the left eq.\eqref{eq:BetheRescale} by summing both sides of 
that equation over all $j$.

As opposed to the  XXX$_{+1/2}$ case,
we expect the roots for the ground state to lie on the real axis
(this may be verified by explicit solution of the exact
Bethe equations for small values of $J$). Furthermore, 
we assume the distribution of roots to be symmetric w.r.t.~the imaginary
axis, $\rho(-u)=\rho(u)$. 
We therefore expect the support of the root density to 
split into (at least) two disjoint intervals 
$\mathcal{C}=\mathcal{C}^- + \mathcal{C}^+$ with
$\mathcal{C}^- =[-b,-a]$ and $\mathcal{C}^+ =[a,b]$,
where $a<b$ are both real.%
\footnote{
After  the analytical continuation to the spin $+\frac{1}{2}$ case, 
the points $a,b$ become a complex conjugate pair.}
For the ground state we expect
just two contours, and the mode numbers should be 
$n=\mp 1$ on $\mathcal{C}^{\pm}$.
For this distribution of roots, the Bethe equations
\eqref{eq:BetheCont} become
\[\label{eq:airfoil}
\pint_a^b \frac{dv\,\rho(v)\,u^2}{v^2-u^2}
=\frac{1}{4}-\frac{\pi}{2}\,u\ , 
\qquad
\delta_1=\frac{1}{4\pi^2}\int_a^b\frac{du\,\rho(u)}{u^2} \ .
\]
Comparing to the previous solution
in \cite{Beisert:2003xu}, we thus find an identical equation except
that the new filling fraction $\beta=\frac{S}{J}$ is related to the 
previous one 
$\alpha=\frac{J_2}{J}$ by $\beta\to-\alpha$!
\footnote{To 
facilitate comparison,  note that here we are using a different
convention for normalizing the density.}
Interestingly, we already analyzed the case 
of negative $\alpha$, i.e.~positive $\beta$ as a technical trick in 
\cite{Beisert:2003xu}; here we find that this case, which did not
previously correspond to a physical situation for the spin
$+\frac{1}{2}$ chain, is {\it physical} in the case of the 
$-\frac{1}{2}$ chain. The solution of the integral equation
(see, e.g.,~\cite{Kostov:1992pn,Muskhelishvili}),
yielding the density $\rho(u)$, may be obtained
explicitly (in \cite{Beisert:2003xu} we rather eliminated
the density after obtaining an integral representation for it); it
reads
\[\label{eq:denssol}
\rho(u)=\frac{2 }{\pi u} \pint_a^b  
\frac{dv\, v^2}{v^2-u^2}
\sqrt{\frac{(b^2-u^2)(u^2-a^2)}{(b^2-v^2)(v^2-a^2)}}\,.
\] 
This density may be expressed explicitly through standard functions: 
\[\label{eq:bigpi}
\rho(u)=\frac{1}{2 \pi b u} 
\sqrt{\frac{u^2-a^2}{b^2-u^2}}
\left[\frac{b^2}{a}-4 u^2 \Pi\left(\frac{b^2-u^2}{b^2},q\right) \right],
\qquad
q=\frac{b^2-a^2}{b^2},
\]
where we introduced the modulus $q$,
playing the role of an auxiliary parameter, and
$\Pi$ is the elliptic integral of the third kind:
\[\label{eq:third}
\Pi(m^2,q)\equiv 
\int_0^{\pi/2}\frac{d\varphi}{(1-m^2 \sin^2 \varphi)
\sqrt{1-q\sin^2 \varphi}}\ .
\]
Furthermore, we may derive two conditions determining the
interval boundaries $a,b$ as a function of the filling fraction
$\beta$:
\[\label{eq:conds1}
\int_a^b   \frac{ du\ u^2}{\sqrt{(b^2-u^2)(u^2-a^2)}}=\frac{1+2 \beta}{4}
\qquad {\rm and} \qquad
\int_a^b \frac{du}{\sqrt{(b^2-u^2)(u^2-a^2)}}=\frac{1}{4 a b}.
\]
The first is derived from the normalization condition 
eq.\eqref{eq:norm}, while the second is a consistency condition,
assuring the positivity of the density.
These may be reexpressed through standard elliptic 
integrals of, respectively, the second and the first kind; one finds
\[\label{eq:conds2}
\ellE(q)\equiv \int_0^{\pi/2} d\varphi\ \sqrt{1-q\sin^2 \varphi}\,
= \frac{1+2 \beta}{4 b},
\qquad
\ellK(q)\equiv \int_0^{\pi/2}\frac{d\varphi}{\sqrt{1-q\sin^2 \varphi}}
=\frac{1}{4 a}
\]
It is straightforward to eliminate the interval boundaries $a,b$
from these equations; furthermore,  we can integrate the
density and 
compute the energy $\delta_1$ from the right equation in
eqs.\eqref{eq:airfoil}
(cf. \eqref{eq:gaugeembryo})
\[\label{eq:gaugebaby}
\delta_1=\frac{1 }{2 \pi^2} \,
\ellK(q)\bigbrk{(2-q)\ellK(q)-2\ellE(q)} 
,\qquad\ \ \ 
\beta\equiv  \frac{S}{J} 
=\frac{1}{2\sqrt{1-q}}\,\frac{\ellE(q)}{\ellK(q)}-\frac{1}{2}  \ . 
\]
Finally, we can also express the boundaries of the Bethe strings
through the modulus $q$:
\[\label{eq:ab}
a=\frac{1}{4 \ellK(q)},\qquad \ \ \ 
b=\frac{1}{4 \sqrt{1-q}\,\ellK(q)}.
\]
This completes the solution.

\section{The circular vs. imaginary solution}
\label{app:E}

In \cite{Beisert:2003xu} a solution different from the 
type discussed in Appendix~\ref{app:C} was found. 
The resulting anomalous dimension matched the energy of a 
circular string \cite{Frolov:2003qc} at one point of the parameter space,
$J_2=J_1=J/2$. 
Recently the circular string solution was extended to 
all values of $J_2$ \cite{Arutyunov:2003uj} where it was
also shown that the agreement with gauge theory persists 
up to a few orders in a perturbative expansion around
$J_2-J/2$.
Here, we will complete the analysis and prove the correspondence
at the analytic level. We are grateful to Gleb Arutyunov 
for his collaboration on this Appendix.
Without further details of the derivation, we present 
the final results starting with gauge theory. 

There are two conditions on 
the endpoints $is,it$, $s<t$, of the Bethe strings
that arise in the solution \cite{Beisert:2003xu}
($\alpha=J_2/J$):
\[\label{eq:IMconds1}
\int_{-s}^s \frac{dv\ v^2}{\sqrt{(s^2-v^2)(t^2-v^2)}}=\frac{1-2\alpha}{4}
\qquad {\rm and} \qquad
\int_{-s}^s \frac{dv}{\sqrt{(s^2-v^2)(t^2-v^2)}}=\frac{1}{4 st}.
\]
Notice the great similarity to \eqref{eq:conds1}!
We perform the elliptic integrals and get
\[\label{eq:IMconds2}
\ellK(r)-\ellE(r)= \frac{1-2\alpha}{8 t},
\qquad
\ellK(r)=\frac{1}{8 s},
\qquad
r=\frac{s^2}{t^2}.
\]
The differences to \eqref{eq:conds2} are due to the
different regions of integration.
Solving for $s,t$ and substituting in the expression for the energy 
(we use the notation $\hat\delta$ and $\hat \epsilon$ 
to distinguish the circular solution)
\[
\hat\delta_1=\frac{1}{32\pi^2}\lrbrk{\frac{1}{s^2}+\frac{1}{t^2}-\frac{2(1-2\alpha)}{st}}
\]
we get the one-loop result from gauge theory
\[\label{eq:IMgauge}
\hat\delta_1=\frac{2}{\pi^2}\,\ellK(r)\bigbrk{2\ellE(r)-(1-r)\ellK(r)},\qquad
\alpha\equiv \frac{J_2}{J}=\frac{\sqrt{r}-1}{2\sqrt{r}}+\frac{1}{2\sqrt{r}}\,\frac{\ellE(r)}{\ellK(r)}.
\]

The circular string is obtained by the same ansatz \eqref{jj} as for the
folded string (see Appendices~\ref{app:A},\ref{app:B}).
The only difference is that the function 
$\psi(\sigma)$ is now assumed to be periodic modulo
$2\pi$
\[
\psi(\sigma+2\pi)=\psi(\sigma)+2\pi.
\]
Instead of folding back into itself, the
string wraps completely around a great circle.
The set of equations that
describes this circular string are \cite{Arutyunov:2003uj}
\[\arraycolsep0pt\begin{array}{rclcrcl}
\mathcal{J}_2\eq\displaystyle\frac{\wup_2}{y}\lrbrk{1-\frac{\ellE(y)}{\ellK(y)}},&\qquad&
\mathcal{J}_1\eq\displaystyle\frac{\wup_1}{y}\lrbrk{y-1+\frac{\ellE(y)}{\ellK(y)}},
\\[15pt]
\mathcal{E}^2\eq\displaystyle\wup_1^2+\frac{\wup_2^2-\wup_1^2}{y},&\qquad&
\ellK(y)\eq\displaystyle\frac{\pi}{2}\sqrt{\frac{\wup_2^2-\wup_1^2}{y}}.
\end{array}
\]
When solved for $\wup_1,\wup_2$ we get a system of two equations
similar to the one in \eqref{eq:StringS5}
\<
\lrbrk{\frac{\mathcal{E}}{\ellK(y)}}^2-
\lrbrk{\frac{y\mathcal{J}_1}{(1-y)\ellK(y)-\ellE(y)}}^2
\eq\frac{4}{\pi^2},
\nln
\lrbrk{\frac{y\mathcal{J}_2}{\ellK(y)-\ellE(y)}}^2-
\lrbrk{\frac{y\mathcal{J}_1}{(1-y)\ellK(y)-\ellE(y)}}^2
\eq\frac{4}{\pi^2}\,y.
\>
The ansatz for the circular solution is symmetric under 
$\mathcal{J}_1\leftrightarrow\mathcal{J}_2$, but superficially 
this does not seem to apply to these equations.
Indeed, a modular transformation is required to 
interchange $\mathcal{J}_1,\mathcal{J}_2$:
\[
\ellK(y)=\sqrt{1-y'}\,\,\ellK(y'),\qquad
\ellE(y)=\frac{\ellE(y')}{\sqrt{1-y'}},\qquad
y=1-\frac{1}{1-y'}.
\]
In order to make contact with gauge theory, we set 
$y=y_0+y_1/\mathcal{J}^2+\ldots$ and expand the energy 
in powers of $1/\mathcal{J}^2$. Using the expansion
\eqref{eq:expa} we find
\[\label{eq:IMstring}
\hat\epsilon_1=\frac{2}{\pi^2}\,\ellK(y_0)\ellE(y_0),\qquad
\alpha\equiv\frac{J_2}{J}=\frac{1}{y_0}-\frac{\ellE(y_0)}{y_0\,\ellK(y_0)}.
\]
As before, the string solution \eqref{eq:IMstring}
is related to the gauge solution \eqref{eq:IMgauge} through a modular transformation
\[
\ellK(y_0)=(1-\sqrt{r})\ellK(r),\qquad
\ellE(y_0)=2(1-\sqrt{r})^{-1} \ellE(r)-(1+\sqrt{r})\ellK(r)
\]
where 
\[\label{eq:IMmap}
y_0=-\frac{4\sqrt{r}}{(1-\sqrt{r})^2}=-\frac{4ab}{(b-a)^2}.
\]
Note that the integration constant of the circular string 
$y$ is related to the integration constant of the folded string
$x$ by $y=1/x$.
The gauge theory constants $a=is,b=it$ describe the endpoints
of some Bethe strings.
Remarkably, \eqref{eq:IMmap} is exactly 
the same relation as in the case of the folded string \eqref{eq:map}!

\section{Energy as a function of the spins}
\label{app:D}

In this Appendix we shall 
discuss the behavior of the leading term 
in the classical energy for the two-spin 
string solutions in  different regions of the 
parameter space $J_2/J$ or $S/J$, respectively.
\begin{figure}\centering
\includegraphics{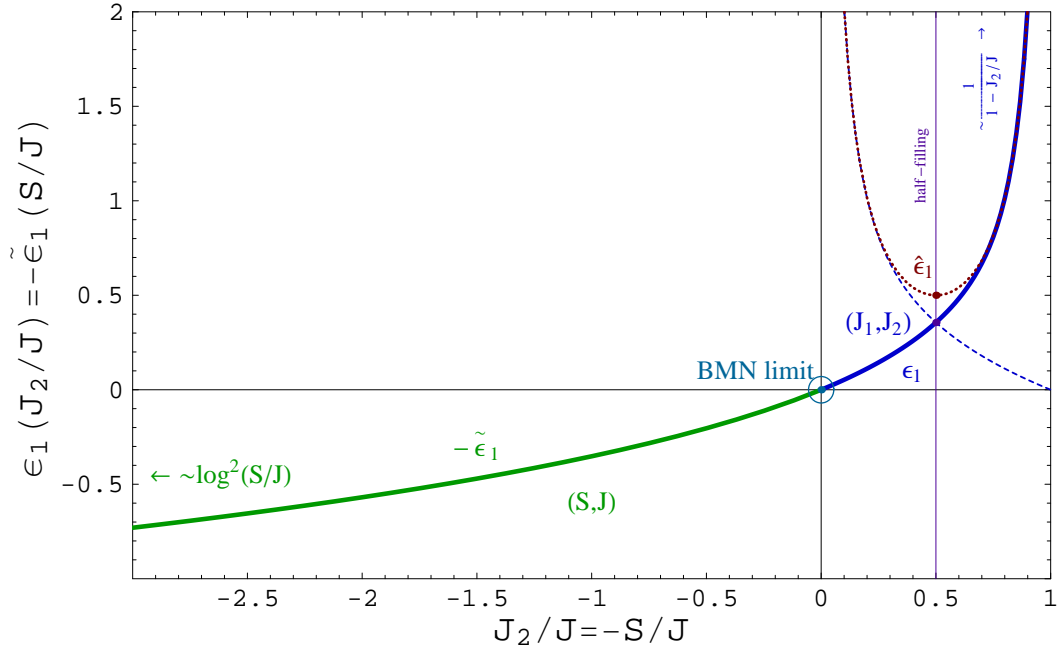}\qquad\qquad\phantom{}
\caption{
The leading order correction to the energy $\epsilon_1$ 
for the folded string solution.
The region $J_2/J>0$ correspond to the $(J_1,J_2)$ case,
whereas $J_2/J<0$ correspond to the $(S,J)$ case with
$S/J=-J_2/J$ and energy $\tilde \epsilon_1=-\epsilon_1$.
We also plot a mirror image under the symmetry
$J_1\leftrightarrow J_2$ (dashed) and the 
energy of the circular string $\hat\epsilon_1$ (dotted).}
\label{fig:plot}
\end{figure}
For the  $(J_1,J_2)$ solution  the function 
$\epsilon_1(J_2/J)$ is defined in the 
region $0\leq J_2/J<1$
whereas for the $(S,J)$ solution $\tilde \ep_1(S/J)$ 
is naturally defined for 
$0\leq S/J< \infty$.
When analytically continued, 
these functions are related by 
$\epsilon_1(j)=-\tilde \epsilon_1(-j)$.
In Figure~\ref{fig:plot}, we 
therefore plot the function $\epsilon_1(j)$ in the
region $j\in (-\infty,1)$.

Depending on the region of the parameter space  
one finds different functional 
form of dependence of the energy on the two spins. 
For example, in the case of the $(S,J)$ solution there are two 
different asymptotics that were considered in \cite{Frolov:2002av}:
The string can become very long and 
approach the boundary of $AdS_5$, i.e.~$\rho_0\to\infty$; 
the energy of this configuration is
\[
E=S+J+\frac{\lambda}{2 \pi^2 J}\, \ln^2 \frac{S}{J} + \ldots
\,, \qquad\ \ \ \ \   S\gg J  \,.
\]
In gauge theory this corresponds to the case where
the two contours $\mathcal{C}^\pm$ meet, i.e.~$a\to 0,b\to \infty$.
A short string reproduces the energy of BMN operators correctly%
\footnote{The BMN case corresponds to 
expanding near a point-like string moving along big circle of $S^5$.
In the limit $J \to \infty,\  \frac{\lambda}{J^2}$=fixed one 
may drop all but quadratic fluctuation terms in the string action
(which becomes then equivalent to the plane-wave action in the 
light-cone gauge). The energies of fluctuations above the BPS 
ground state $E=J$ are then determined by the string 
fluctuation masses given  
by  $m^2= \frac{1}{\cJ^2}=
\frac{\lambda}{J^2}$.}
(all mode numbers are $n=\pm 1$)
\[
E=J + S + \frac{\lambda\,S}{2J^2}+\ldots\ ,\qquad \ \ \ \  S \ll J \,.
\]
One may also consider a ``near BMN'' limit $S/J\ll 1$ of the 
two-spin solution keeping full dependence on $\lambda$.
Then we get 
\[
E=J+S \sqrt{1+\frac{\lambda}{J^2}}
-\frac{\lambda\,S^2}{4J^3}\cdot \frac{2+\lambda/J^2}{1+\lambda/J^2}
+\ldots \ , \ \ \ \ \  \ \   S \ll J \ ,  
\]
This gives the near BMN limit for a total of $S$ excitations of modes $n=\pm 1$. 
Note, however, that we must consider a large number of 
excitations $S=\order{J}$, i.e. $S=\beta J$ with 
$\beta$ small but $\order{J^0}$.
Therefore, we may \emph{not} assume that $S=\beta J$ takes a particular, 
finite value like, for example, two (in an attempt to 
compare to $1/J$ terms in \cite{Callan:2003xr}).
Instead, one must consider an arbitrary number of excitations $S$ 
and consider only the coefficient $c$ in the near BMN correction
$(cS+c')/J=c\beta+\order{1/J}$. 
The point is that  $c'$ is an 
$\order{1/J}$ correction which we  presently ignore.

At $S/J=0$ we make the ``Wick-rotation'' to the $(J_1,J_2)$ solution.
The energy of a short string rotating on $S^5$ is given by
\[
E=J+
\frac{\lambda\,J_2}{2J^2}+\ldots \,,\qquad \ \  J_2 \ll J   \,,
\]
where we can explicitly see the connection to the $(S,J)$ case.
Similarly, the near BMN limit reads
(we set $J_1=J$ to compare to BMN terminology,
$J_2$ represents the number of excitations)
\[
E=
J_1
+J_2 \sqrt{1+\frac{\lambda}{J_1^2}}
-\frac{\lambda\,J_2^2}{4J_1^3}\cdot \frac{2+3\lambda/J_1^2}{1+\lambda/J_1^2}
+\ldots \ , \ \ \ \ \ \    J_2 \ll J \ .  
\]
As the charge $J_2$ increases, the string grows until for
$J_1=J_2$ we get
\[\label{eq:half}
E=J+ c \frac{\lambda}{J}+\ldots
 \ , \ \ \  c \equiv \ep_1 (1/2) = 0.356016\ldots 
 \ , \ \ \ \ \  J_2=J_1=J/2 \ . 
\]
In string theory nothing special happens,
the string extends over approximately $120^\circ$ and can grow further.
In contrast, in gauge theory we made the assumption $J_2\leq J_1$ to 
solve the Bethe ansatz. Therefore,  we do not get solutions beyond
this point.
Nevertheless, in terms of charges, we can freely interchange $J_1$ and $J_2$. 
The string solutions for $J_2>J/2$ should correspond to some
gauge theory states with $J_2<J/2$
\[
\delta'_1(J_2/J)=\epsilon_1(1-J_2/J)\ ,\ \ \ 
\qquad\mbox{for }\ 0\leq J_2 \leq J/2\ .
\]
We see that the string energy $\epsilon_1(J_1,J_2)$ is not symmetric 
with respect to $J_1\leftrightarrow J_2$.
As a consequence, 
 the anomalous dimensions $\delta'_1(J_1,J_2)$ 
do not belong to operators with the
minimal energy, $\delta_1(J_1,J_2)$, but to some other
set of operators with larger dimensions.
That suggests  that one and the same string solution
describes two different operators in different regions of parameter space.
There are some indications%
\footnote{
Apparently, solutions to the Bethe equations 
with $J_2>J/2$ correspond to mirror images 
of solutions with $J'_2=J+1-J_2\leq J/2$
($s\mapsto -1-s$ with $SU(2)$ spin $s=J/2-J_2$).
If we assume $J_2$ and $J$ to be even,
in this way we would find solutions
with odd $J'_2$ and even $J$,
i.e.~the odd, unpaired ground states.}
that these new gauge theory operators
are the odd, unpaired ground state solutions 
found in \cite{Beisert:2003xu}.
This is an interesting possibility, as it would explain why 
for half-filling, $J_2=J/2$, the odd, unpaired ground state has
energy \eqref{eq:half} as suggested by numerical 
evidence \cite{Beisert:2003xu}.
Further numerical evidence shows that the anomalous dimension
 $\delta'_1$ of 
this state near $J_2=0$ scales as $1/J_2$ instead of $1/J$. 
Indeed,  this is what happens on the string theory side.
The largest extension of the $S^5$ solution takes place near $J_2=J$.
Then the string extends over half a great circle and the energy is
\footnote{Since the one-loop correction $\epsilon_1$
grows to infinity at  $J_2\approx J$ 
the string solution is not stable at large enough $J_2$.}
\[
E=J+\frac{2\,\lambda}{\pi^2\,J(1-J_2/J)}+\ldots=J+\frac{2\,\lambda}{\pi^2\,J_1}+\ldots
 \ , \ \ \ \ \ \    J_2\approx J \ .  
\]
At $J_2=J$ the folded string becomes, in fact, 
equivalent to a different configuration:
One half of the string can be 
unfolded to give the circular string
discussed in Appendix~\ref{app:E}.
It is interesting to see that also the energy of the 
circular solution $\hat\epsilon_1$ asymptotes to the same value
\[
E=J+\frac{2\,\lambda}{\pi^2\,J_1}+\ldots
 \ , \ \ \ \ \ \    J_2\approx J \ .  
\]
The energy $\hat\epsilon_1$ of the circular solution 
decreases as we decrease $J_2$ up to half-filling
$J_2=J_1=J$. Unlike in the case of the folded string,
the energy has a minimum 
\[
E=J+\frac{\lambda}{2J}+\ldots
 \ , \ \ \ \ \ \    J_2=J_1=J/2 \ .  
\]
Furthermore, the solution $\hat\epsilon_1$ is symmetric 
under $J_1\leftrightarrow J_2$ and we can stop.


\bibliography{compare}
\bibliographystyle{nb}

\end{document}